\documentclass[aps,prb,twocolumn,epsfig,showpacs]{revtex4}
\usepackage{graphicx} 
\begin{document}

\title{Slow dynamics of a confined supercooled binary mixture: 
direct space analysis}

\author{P.~Gallo$^\dagger$\footnote[1]{Author to whom correspondence 
should be addressed; e-mail: gallop@fis.uniroma3.it
}, R.~Pellarin $^\dagger$ and M.~Rovere$^\dagger$,}
\address{$\dagger$ Dipartimento di Fisica, 
Universit\`a ``Roma Tre'', \\ Istituto Nazionale per la Fisica della Materia,
 Unit\`a di Ricerca Roma Tre\\
Via della Vasca Navale 84, 00146 Roma, Italy.}

\begin{abstract}
Dynamical properties of a Lennard-Jones 
binary mixture embedded in an off lattice matrix of soft spheres 
are studied in the direct space
upon supercooling by molecular dynamics simulations.
On lowering temperature the smaller particles tend to
avoid the soft sphere interfaces and correspondingly their
mobility decreases below the one of the larger particles.
The system displays a dynamic behaviour consistent with the Mode Coupling 
predictions. 
A decrease of the mode coupling crossover temperature with respect to
the bulk is found.
We find however that the range of validity of the theory shrinks
with respect to the bulk. This is due to the change in the 
smaller particle mobility and to a substantial 
enhancement of hopping processes well above the cross over temperature
upon confinement.

\end{abstract}

%%%%%%%
%61.20.Ja Computer Simulation of Liquid Structure
%61.20.Lc Time dependent properties, relaxations
%64.70.Pf Glass transition
%\date{\today}
%%%%
\maketitle

\section{ Introduction}

Liquids under confinement represent a field of growing interest
in science due to the connection with relevant technological
and biophysical problems~\cite{grenoble}. In this field modifications of 
both the phase diagram and dynamics
of the confined liquid with respect to the bulk represent a key point,
in particular as far as the possibility of supercooling 
is concerned~\cite{pablo}.
Experiments exhibit a very diversified phenomenology 
for supercooled confined liquids.
In particular the calorimetric glass transition temperature, $T_g$,
can increase~\cite{melni}, decrease~\cite{pissis,kremer2} 
or be unaffected~\cite{richert} with respect to the the bulk value 
depending on the liquid, the confining geometry and the 
nature of the substrate.

Phenomenological arguments predict the existence 
of domains of cooperative dynamics in the supercooled 
liquid~\cite{adam-gibbs}.
The size of these domains is expected
to grow upon supercooling. 
The existence of an upper bound
for the domain size in confined media
implies a decrease of $T_g$ upon confinement.
In fact if this size remains small on supercooling then the probability of
a cooperative motion, and therefore of a structural relaxation, is large,
and dynamics is faster.

The mode coupling theory of the glass transition (MCT)~\cite{goetze} is 
able to describe the dynamics of bulk liquids 
in the supercooled region on approaching a crossover temperature
$T_C$. At $T_C$ the system passes 
from a regime where ergodicity is 
attained through structural relaxations
to a regime where this mechanism is frozen and
only activated processes permit the exploration 
of the configurational space. 
$T_C$ can be estimated through experiments and computer
simulations or predicted by the theory.

Above $T_C$
the relaxation mechanism of the supercooled liquid can be described
as mastered by the 
cage effect. Nearest neighbors surround and trap the tagged particle
forming a cage around it. When the cage relaxes, 
due to cooperative motions, the particle moves.
MCT translates this 
phenomenological description into a precise mathematical
framework where, starting from the classical equation of motion for
the density correlator $\phi_q(t)$, a retarded memory function is introduced.
In the idealized version of MCT hopping processes are neglected 
and the retarded non linear set of integro-differential equations can
be solved analytically to the leading order in $\epsilon=(T-T_C)/T_C$,
deriving universal results  for the density correlator.
With these approximations $T_C$ is the temperature of structural arrest
of the ideal system. The success of this theory is due to the fact
that on approaching $T_C$ from above hopping processes can be neglected in 
many liquids and the predictions of the idealized version of MCT hold. 

Molecular dynamics studies of model liquids
in restricted geometries intended to assess the applicability
of MCT can give an important contribution 
to the characterization of the glass
transition scenario in confinement.
In this field recent progresses have been done for water confined
in a silica nano pore~\cite{gallo1,gallo2} 
and for confined thin polymer films~\cite{baschnagel}.

We performed a Molecular Dynamic (MD)
study of the single particle dynamical properties
of a Lennard-Jones binary mixture (LJBM) embedded in an off
lattice matrix of soft spheres. 
The model considered represents a situation
of strong confinement analogous to real silica 
xerogels~\cite{rosinberg,monson,page}. 
The bulk phase of the chosen mixture is known to test most of
the MCT predicted features~\cite{kob}.
A brief report on some of our
findings has been recently published \cite{euro-noi}. 
 
We present in this paper direct space quantities  
evaluated from MD trajectories in order to define
the range of validity of MCT with respect to the bulk phase. 
In the next section we describe the model adopted and the
details of the MD simulations. In the 
third section we illustrate the radial pair distribution functions
and the coordination numbers of the system. 
The fourth section is devoted to the study of
the dynamic quantities in the direct space:
the mean square displacement (MSD), 
the van Hove self correlation function
and the non-Gaussian parameter. 
Last section contains the concluding remarks.

%%%%%%%%%%%%%%%%%%%%%%%%%%%%%%%%%%%%%%%%%%%%%%%%%%%%%%%%%%%%%%%%
\begin{figure}[ht]
\includegraphics[width=8cm]{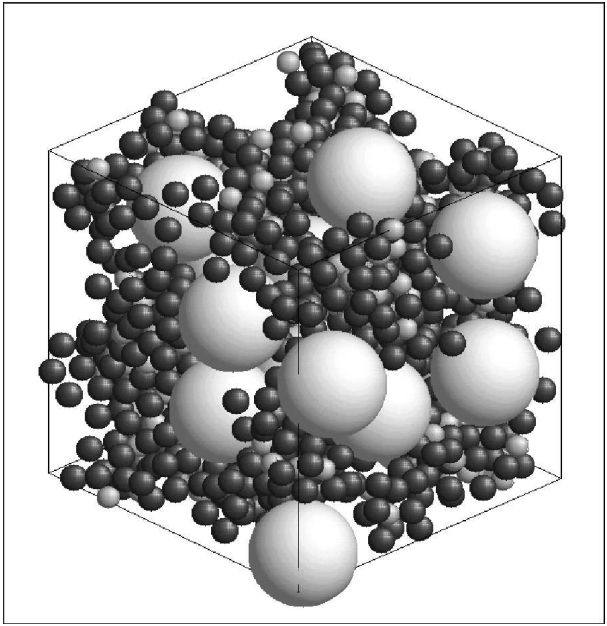}
\caption{Graphic representation of the model studied. Small light
gray particles are of A type, small dark
gray particles are of B type. The soft spheres that form the confining
system appear in the figure as large dark gray spheres.
The cube in which the system is embedded is the simulation box.
The radii of the three types of particles are purely representative.
The $\sigma$ values are listed in Tab.\ref{tab1}.}
\protect\label{fig:1}
\end{figure} 
%_______________________________________________________________
%%%%%%%%%%%%%%%%%%%%%%%%%%%%%%%%%%%%%%%%%%%%%%%%%%%%%%%%%%%%%%%%
\begin{figure}[ht]
\includegraphics [width=7cm]{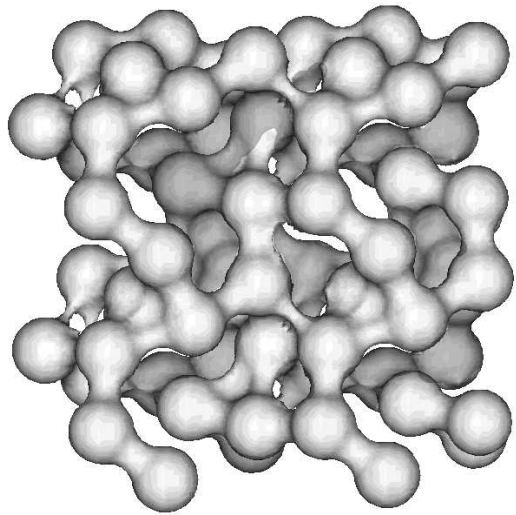}
\caption{Graphic representation of the confining disordered structure. 
The sixteen spheres of the matrix contained in the
simulation box of Fig.\protect\ref{fig:1}
have been repeated to obtain a box of height 2L,
width $2L$ and depth $L$. The surface that appears in the 
figure is equipotential.}
\protect\label{fig:2}
\end{figure} 
%_______________________________________________________________

\section{Model and Molecular dynamics}

The confining medium consists in a rigid disordered array of soft 
spheres. The simulation box contains 16 soft spheres labeled
in the following with the letter M. 
The interspaces of this structure host a 
liquid LJBM 
of 1000 particles, composed of 800 particles 
of type A and 200 particles of type B. 
The interaction
pair potential can be written in the following general form: 
\begin{equation}
V_{\mu\nu}(r)=4\epsilon_{\mu\nu}\left[\left(\frac{\sigma_{\mu\nu}}
{r}\right)^{12}-\eta_{\mu\nu}\left(\frac{\sigma_{\mu\nu}}
{r}\right)^{6}\right]
\label{potential}
\end{equation}
where indexes $\mu,\nu$ run on the particle types A,B,M. 
The parameters for the three components 
are listed in Table~\ref{tab1}. The parameters of the LJBM 
have been chosen as in ref.~\cite{kob}.
The shifted potential technique has been used with a 
cutoff of $2.5$ $\sigma_{\mu\nu}$.
Periodic boundary conditions are applied.
In the following Lennard-Jones units will be used, therefore
energy will be given in units of
$\epsilon_{AA}$, temperature in units of 
$\epsilon_{AA}/K_B$, length in units $\sigma_{AA}$ and time in units of
$(m\sigma_{AA}^2/(48\epsilon_{AA}))^{1/2}$.
The box length is $L=12.6$. 
The simulation box of the system is 
represented in Fig.~\ref{fig:1}. 
As it can be seen from the picture
the LJBM is in a strongly confined 
environment as only few layers of particles can accommodate 
among the soft spheres. 
Fig.~\ref{fig:2} shows a cell made of four simulation boxes.
Only equipotential surfaces of the soft spheres are 
displayed in 
order to offer a clearer image of
the disordered porous structure hosting the liquid
and to highlight the analogies with silica xerogels.

We have conducted MD simulations of this system in the NVE ensemble.
The equations of motion were solved by the velocity Verlet algorithm.
The system was equilibrated at different reduced temperatures 
via a velocity rescaling procedure.
The temperatures investigated are 
$T=5.0,2.0,0.8,0.58,0.538,0.48,0.465,0.43,0.41,0.39$.
The timestep used for $T\ge 1$ was $0.01$ and  for  $T<1$ was $0.02$. 
For the lowest temperature investigated, $T=0.37$, a production run of 
$14\times 10^6$ timesteps was performed. 
We verified that the results obtained
do not depend on the specific choice of the disordered matrix
by running MD simulations for two other different configurations
of the disordered matrix of soft spheres.
Both thermodynamics and dynamics appeared the same for the three systems, 
in particular deviations among the diffusion coefficients 
extracted at given temperatures from the MSD
of different configurations are within $2$\%.

\begin{figure}[ht]
\includegraphics[width=8cm] {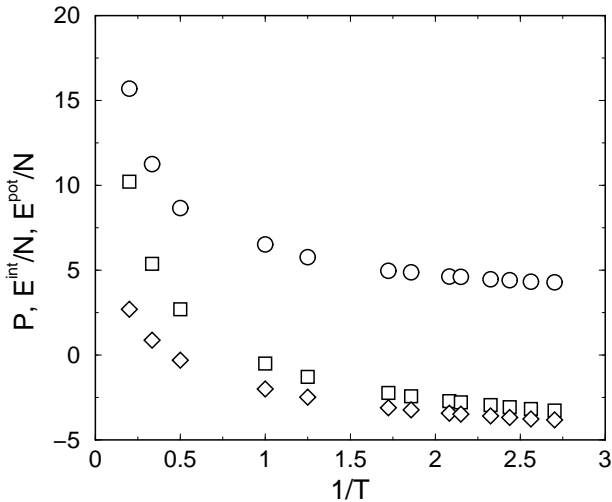}
\caption{Thermodynamics of the confined Lennard-Jones binary mixture:
pressure (circles), total energy per particle (squares) and potential
energy per particle (diamonds) versus the inverse of the
temperature.}
\protect\label{fig:3}
\end{figure}

Values of thermodynamics quantities of the equilibrated systems are
shown in Fig~\ref{fig:3}. Averaged pressure, potential energy 
and total energy per particle of the mixture behave similar to the bulk.
A stability of the system throughout all the isochore explored is evident
from the smoothness of the curves.

\begin{figure}[ht]
\includegraphics[width=8cm]{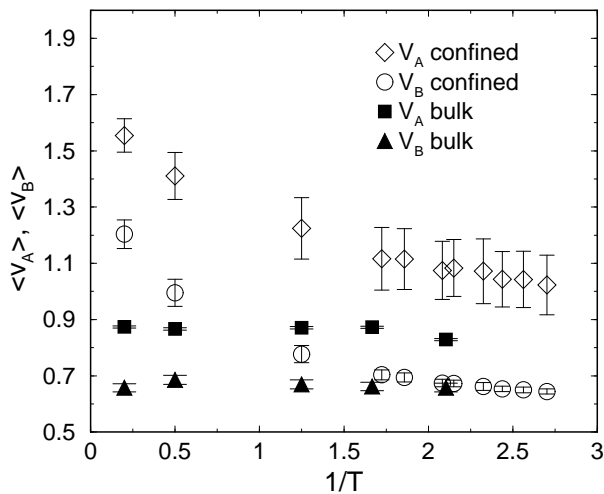}
\caption{Specific volumes of the Voronoi cells for A and B particles.
Particles in contact with the soft confining spheres
have been excluded. The bulk considered is that of ref.~\cite{kob}.
}
\protect\label{fig:7a}
\end{figure}

Due to the
soft spheres potential the free volume accessible for the mixture
strongly depends on temperature.
In Fig.~\ref{fig:7a} we show an estimate of the free volume for
A and B particles in our confined system as a function of temperature.
The estimate has been carried out through the Voronoi tessellation.
This procedure has been applied only to non-interfacial particles.
The free volume has been calculated by averaging over $160$ configurations
of the system for each temperature.

%%%%%%%%%%%%%%%%%%%%%%%%%%%%%%%%%%%%%%%%%%%%%%%%%%%%%%%%%%%%%%%%
%_______________________________________________________________
%%%%%%%%%%%%%%%%%%%%%%%%%%%%%%%%%%%%%%%%%%%%%%%%%%%%%%%%%%%%%%%%

%_______________________________________________________________

\section{Radial pair correlation functions}

\begin{figure}[ht]
\includegraphics[width=7cm]{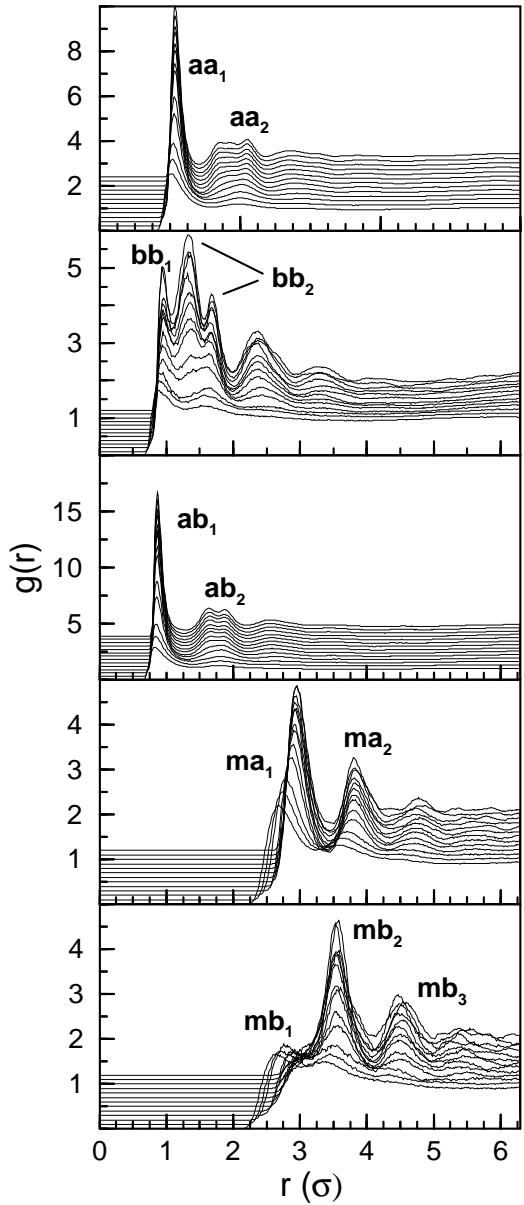}
\caption{Radial distribution functions 
calculated at different temperatures for the system.
From top to bottom AA, BB, AB, MA and MB. Lower temperatures are
progressively shifted upward. For the AA function the shift is
0.2, for AB 0.3 while for the remaining BB, MA and MB it is 0.1.
The labeling of the peaks is used in the text and in 
Fig.\protect\ref{fig:5}.}
\protect\label{fig:4}
\end{figure}

The radial pair correlation functions 
$g_{\mu\nu}(r)$ are shown in Fig.\ref{fig:4}.
For all the temperatures investigated the system is 
in the liquid phase as no abrupt sharpening of the peaks is evident. 
No indications of phase separation appear upon
supercooling either. The
latter phenomenon would in fact result
in a substantial decrease of the area of the first peak, $ab_1$, of the 
$g_{AB}(r)$, which we do not observe in our system.  
On lowering temperature the peaks sharpen and enhance
more than in the bulk due to the strong packing induced by the soft spheres.  
A clear element of distinction
with respect to the bulk is the enhancement of intensity of the first
$g_{BB}(r)$ peak, $bb_1$, as supercooling progresses.
The same peak appeared instead reduced to a small
shoulder for supercooled temperatures in the bulk.

The functions $g_{MA}(r)$ and $g_{MB}(r)$ are characteristic of this 
confined system and are representative of the interaction between the
confining soft spheres and the particles of the mixture.
We note the gradual disappearance of the first peak, $mb_1$, 
of the  $g_{MB}(r)$
that begins at $T=2.0$ while the second peak, $mb_2$, 
enlarges its area. Correspondingly the first peak of the $g_{MA}(r)$
shifts to larger $r$. For lower temperatures, namely $T<0.8$, the structure
of the mixture stabilizes so that MA and MB peaks alternate in space.

\begin{figure}[ht]
\includegraphics[width=8cm]{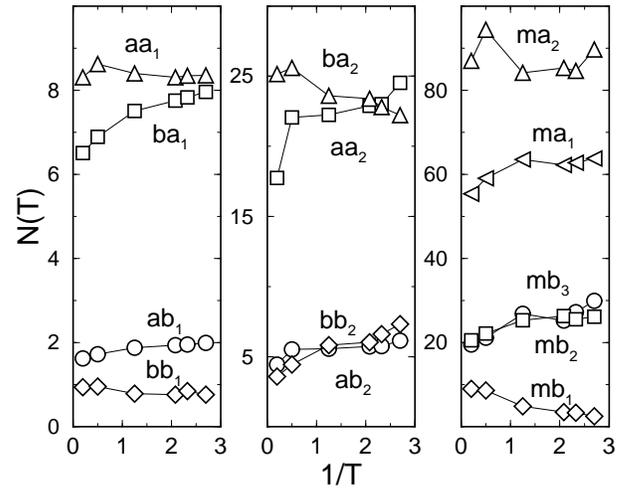}
\caption{Coordination numbers of the shells defined by the peaks
of the radial distribution functions as labeled in 
Fig.\protect\ref{fig:4}. See Eq. \protect\ref{coordnum}.}
\protect\label{fig:5}
\end{figure}

In order
to quantify the above considerations for the radial distribution
functions we calculated the
coordination number defined as:
\begin{equation}
N_{\mu\nu}(r_1,r_2)=4\pi\rho x_\nu \int_{r_1}^{r_2} r^2 g_{\mu\nu}(r) dr 
\label{coordnum}
\end{equation}
where $\rho$ is the total number density $N/V$, where $N=1016$, 
$V=L^3$ and $x_\nu$ is the relative fraction of the $\nu$ species.
The quantity of Eq.~\ref{coordnum}, evaluated for each peak 
of the $g_{\mu\nu}(r)$, is shown in Fig.~\ref{fig:5}. 
The number of B particles around the soft sphere 
in the first shell lowers upon decreasing temperature in favor of an 
enhancement in the second shell. 
Correspondingly the number of A particles in the first shell enhances.

\begin{figure}[ht]
\includegraphics[width=8cm]{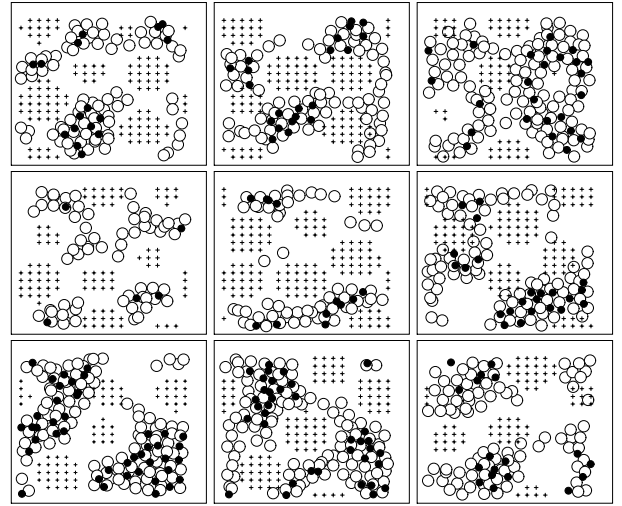}
\caption{Tomographic snapshots of the system at 
the lowest temperature investigated, T=0.37. In order to
obtain this picture the system has been divided in slices of height 
$1.4\sigma$. From top to bottom and from left to right every 
picture displays the content of each slice. Empty circles
are A-type particles, filled circles are B-type. Crosses
mark the volume of the soft spheres. No phase separation is evident.
At this temperature B particles avoid the interfacial region.}
\protect\label{fig:6}
\end{figure}

The above considerations on the radial distribution functions and
the coordination numbers lead to the conclusion that
B particles avoid the interface
with the soft spheres for low temperatures. In Fig.~\ref{fig:6}
a snapshot of the system for $T=0.37$ evidences this behavior.
The fluctuations of the error bars in Fig.\ref{fig:7a}
also confirm this phenomenon.  
In fact since interfacial particles have been excluded from the calculation 
of the free volume a 
larger fluctuation corresponds to a frequent exchange of the particles with
the interface.

%%%%%%%%%%%%%%%%%%%%%%%%%%%%%%%%%%%%%%%%%%%%%%%%%%%%%%%%%%%%%%%%
%_______________________________________________________________
%%%%%%%%%%%%%%%%%%%%%%%%%%%%%%%%%%%%%%%%%%%%%%%%%%%%%%%%%%%%%%%%
%_______________________________________________________________
%%%%%%%%%%%%%%%%%%%%%%%%%%%%%%%%%%%%%%%%%%%%%%%%%%%%%%%%%%%%%%%%
%_______________________________________________________________

\section{Dynamic quantities}

In the following we shall describe the temperature 
variations of the time dependent
direct space quantities analyzed  
for our confined system upon supercooling, namely the MSD,
the van Hove self correlation function and the non-Gaussian parameter.

\begin{figure}
\includegraphics[width=8cm]{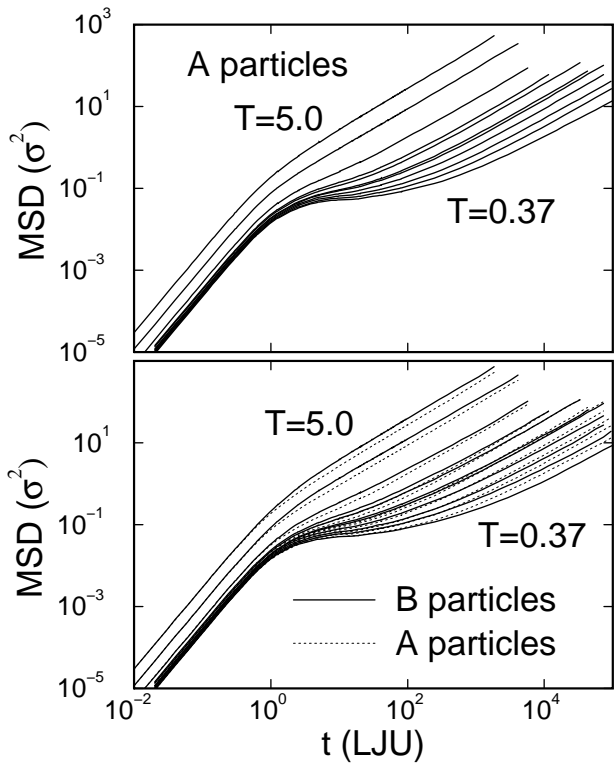}
\caption{Log-log plot of the mean square displacements of
A particles and B particles 
for all the temperatures
investigated, namely 
$T=5.0,2.0,0.8,0.58,0.538,0.48,0.465,0.43,0.41,0.39$.}
\protect\label{fig:7}
\end{figure}

The asymptotic behaviour of the MSD can be used as a 
test of an MCT prediction.
The MSD is in fact expected, as supercooling progresses,
to develop a plateau at intermediate time scale, corresponding to the
rattling of the particle in the cage.
After the initial short time ballistic motion the particle enters the
plateau whose extension in time enhances upon lowering temperature.
After leaving the plateau region the
Brownian behaviour predicted for a simple liquid
in a stable state is restored. From the slope of the late 
MSD the diffusion constant $D$ can be extracted through Einstein relation
$<r^2(t)>=6Dt$.
The temperature behaviour of diffusion coefficients can
give an estimate of the mode coupling crossover
temperature through the MCT power law behaviour
\begin{equation}
D\sim (T-T_C)^\gamma
\label{pwl}
\end{equation}
In Fig.\ref{fig:7} we show the MSD calculated for both A and B particles as
a function of time for all the temperatures investigated.
It can be seen that also in this confined system the binary mixture
displays the typical features of a glass former. 
From the height of the plateau the cage
radius can be extracted and the value obtained for both A and B
particles is $r^2_c=0.04$, analogous to the bulk result~\cite{kob}.

%%%%%%%%%%%%%%%%%%%%%%%%%%%%%%%%%%%%%%%%%%%%%%%%%%%%%%%%%%%%%%%%

%_______________________________________________________________
%%%%%%%%%%%%%%%%%%%%%%%%%%%%%%%%%%%%%%%%%%%%%%%%%%%%%%%%%%%%%%%%
\begin{figure}
\includegraphics[width=8cm]{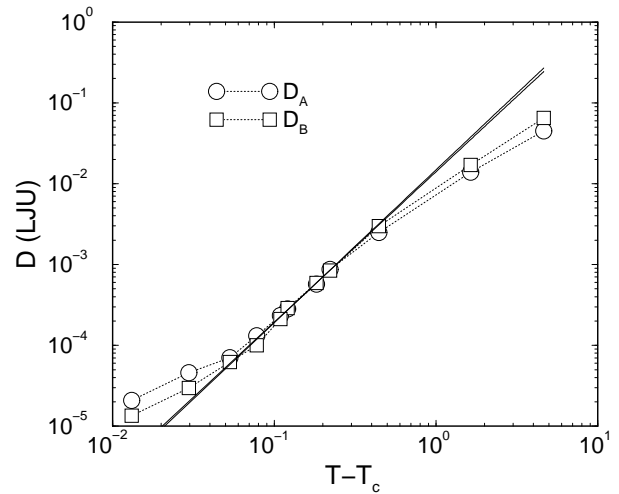}
\caption{Diffusion coefficients for A and B particles 
(symbols). Continuous lines correspond to the fit to Eq. 
\protect\ref{pwl}.
The values of the parameters extracted from the fit are
$T_C=0.356$ for both species and $\gamma=1.86$ for A particles
and $\gamma=1.89$ for B particles.}
\protect\label{fig:8}
\end{figure} 
%_______________________________________________________________

The lowest temperature reached in our study is $T=0.370$ 
against $T=0.466$ for the bulk. The elongation of
the plateau of the MSD for our lowest temperature is comparable to that
of the lowest temperature of the bulk.
The diffusion coefficients extracted from the slopes are reported in 
Fig.~\ref{fig:8}.
At higher temperatures $T>1.0$ the larger A particles diffuse
slower than B particles while they move faster at lower temperatures.
As shown in the previous section, in the confined case 
for lower temperatures B particles
tend to avoid the interface of
the soft spheres. As a consequence they remain confined in the inner part
of the channels, surrounded by A particles, and this diminishes the
accessible free volume lowering mobility below the one of
A particles, see also Fig.\ref{fig:7a}. 

In Fig.~\ref{fig:8}
the fit to the MCT power law of Eq.~\ref{pwl} is also reported.
The estimate of the crossover temperature given by the fit is
$T_C=0.356$ for both A and B particles. The exponents are 
$\gamma=1.86$ and $1.89$ respectively for A and B particles.
We do observe deviation from a power law both at high 
and low temperatures. The temperatures used for the fit 
are in the range $0.410\le T\le0.58$ corresponding to
$0.153<\epsilon<0.631$. 
The range of validity of MCT reported for the bulk
is much wider, $0.07<\epsilon<1.30$. 
The upper limit is generally due in bulk to
the non-validity of the asymptotic
expansion for too large values of $\epsilon$.
Here we infer that a further decrease of this limit is due to the
progressive B particles avoidance of the soft sphere interfaces 
upon supercooling.
For the temperatures below the upper limit 
the peaks of g(r) for MA and MB have in fact already 
stabilized, see Fig.\ref{fig:4}. 
The deviation observed at the two lowest temperatures
is to be connected to the presence of hopping mechanisms,
analogously to bulk liquids,
as we will see in more detail when we illustrate the behaviour of the VHSCF.

The Van Hove self correlation function,
VHSCF, is defined, for a system of N particles, as
\begin{equation}
G_S(r,t)=\frac{1}{N}\langle\sum_{i=1}^N\delta [{\bf{r}}+{\bf{r}}_i(0)-
{\bf{r}}_i(t)]\rangle
\label{grt}
\end{equation}
$4\pi r^2G_S(r,t)dr$ is the probability of finding a particle at distance
$r$ after a time $t$ if the same particle was at the origin 
$r=0$ at the initial time $t=0$.

\begin{figure}[h]
\includegraphics[width=8cm]{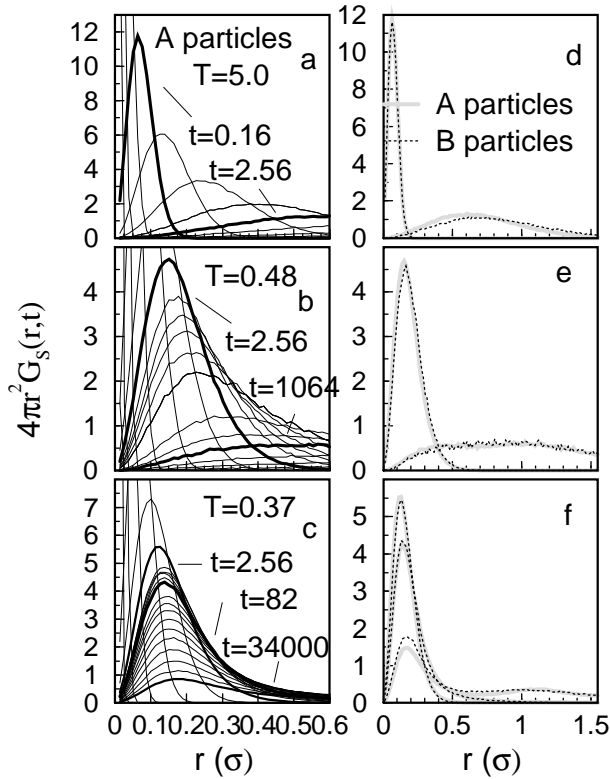}
\caption{$4\pi r^2G_S(r,t)dr$
vs r.
a) b) and c) pictures refer to particle A for T=5.0, 0.5 and 0.48
respectively. Data are sampled at fixed times.
The times chosen follow the progression $t=2^n$.
In the pictures d) e) and f) the correlators are shown
for B particles (dashed lines) together with the correlators 
for A type for a comparison (continuous line). The curves chosen
for a comparison are those indicated with a thicker line in the graphs
on the left.}
\protect\label{fig:10}
\end{figure}

In Fig.\ref{fig:10} we show $4\pi r^2G_S(r,t)dr$ at fixed
times for three different temperatures.
On this scale the correlators of A and B particles do not differ 
substantially.
At high temperature the VHSCF decays regularly as time increases as
expected for a simple liquid. As supercooling progresses 
the system begins to display a new feature that is best evident
for the lowest temperature. This feature consists in the clustering of the 
curves at intermediate times which is the signature of the caging
of the particles. The zone where the peak value clusters is 
$r=0.15-0.25$ and corresponds to the zone in which the MSD flattens. 
For the lowest temperature investigated 
the maximum and minimum times for the clustering are $2.6\le t\le 82.0$.
These values define the so called $\beta$ relaxation temporal region.
In the comparison between the VHSCF of A and B particles 
it can be observed the crossing between the velocity of the two
species. At high temperatures, in fact, the VHSCF of B particles
extends further than the one of A particles for long times, 
while for low temperatures behaviors are swapped.  

%%%%%%%%%%%%%%%%%%%%%%%%%%%%%%%%%%%%%%%%%%%%%%%%%%%%%%%%%%%%%%%%
%_______________________________________________________________
%%%%%%%%%%%%%%%%%%%%%%%%%%%%%%%%%%%%%%%%%%%%%%%%%%%%%%%%%%%%%%%%
\begin{figure}
\includegraphics[width=8cm]{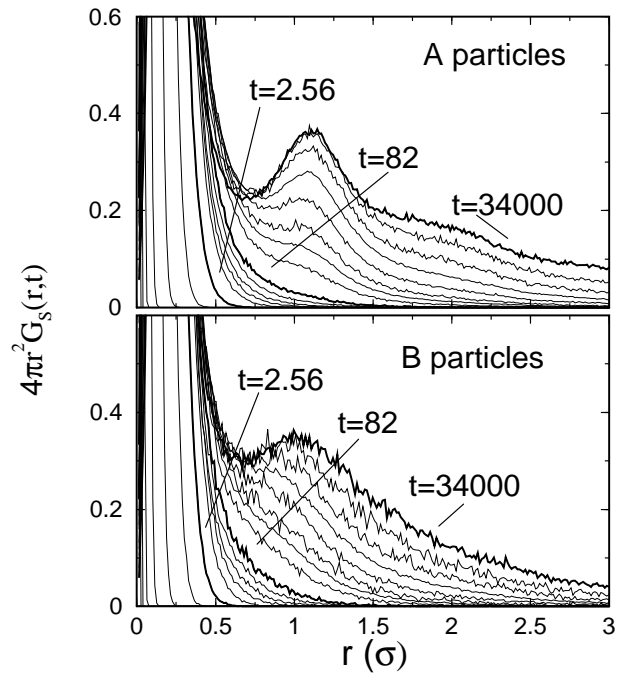}
\caption{Blow up of the $4\pi r^2G_S(r,t)dr$
vs r for T=0.37 for different times for A (top graph) and B (bottom graph) particles.
Evidence is put on the region in which the peaks related to the presence of hopping appear.}
\protect\label{fig:11}
\end{figure} 
%_______________________________________________________________

A behavior not present in the bulk can be observed for the lower
temperatures by looking at the peak of the VHSCF for times
longer than the plateau region.
In this temporal range
the peak of the VHSCF is expected, in the framework of MCT, to shift to
larger $r$ for large times.
The absence of this shift in our
confined mixture, as we shall see in the next picture, 
must be connected with an important hopping phenomenon able to restore 
diffusive motions when the cage is not jet dissolved.

In Fig.\ref{fig:11} we show a blow up of the tails  of the curves
of Fig.\ref{fig:10}. 
Aside the previously cited cage peak, we note the 
presence of a hopping peak around $r=1.0$. This peak 
enhances for late time and it is more pronounced 
for A particles than for B. 
In the bulk on approaching $T_C$ from above a hopping peak appears 
only for B particles.
From the MSD it is evident that
our hopping phenomena can not be neglected 
at the two lowest temperatures, namely at $T=0.37,0.39$. As a 
consequence these temperatures can not be accounted for 
a test of the MCT in the region around the cage relaxation time,
$\alpha$ relaxation region.
Nevertheless 
these temperatures can be used to test the theory 
in the $\beta$-time region. In fact, see also ref.\cite{kob}, hopping peaks 
do not affect curves in the time range of the clustering region.

For the A particle a second broader peak is visible in Fig.\ref{fig:11}.
This peak is also connected to the presence of hopping.
The two hopping peaks correspond to the first and second peak of
the $g_{AA}(r)$, see Fig.\ref{fig:5},
meaning that there are some privileged position for the atoms
to be reached and that those atoms that reached these positions 
have traveled more than the average.
Similar, the hopping peak of the B VHSCF corresponds to the first
peak of the $g_{BB}(r)$.
Hopping phenomena are connected with 
the existence of dynamical heterogeneities
in supercooled liquids \cite{schober}. In confinement they appear
to have above $T_C$ a more important role than in the bulk.
 
Dynamical heterogeneities
are also known to cause a non gaussianity of atomic displacements.
We evaluated the first non-Gaussian parameter (NGP), that works as a quantifier
of the degree of non-Gaussianity.
This parameter is defined \cite{yip} in three dimension as
\begin{equation}
\alpha_2(t)=\frac{3<r^4(t)>}{5<r^2(t)>^2}-1
\end{equation}

%%%%%%%%%%%%%%%%%%%%%%%%%%%%%%%%%%%%%%%%%%%%%%%%%%%%%%%%%%%%%%%%
\begin{figure}
\includegraphics[width=7cm]{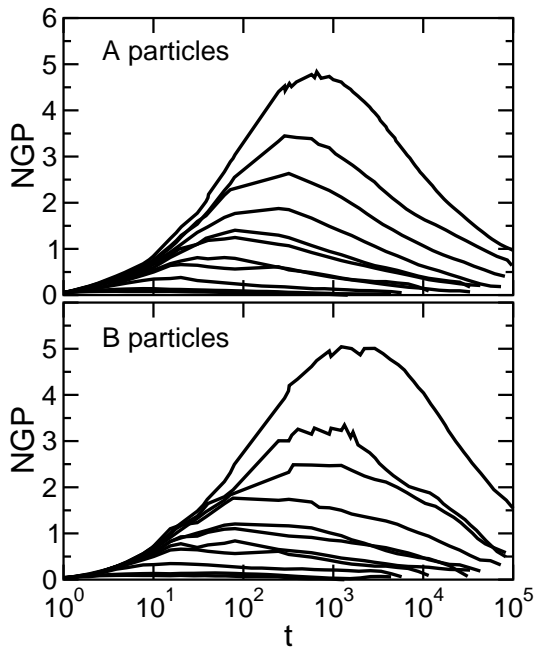}
\caption{Non Gaussian parameter calculated for A 
(top) and B (bottom) particles for all the temperatures investigated.
In both graphs curves on the top correspond to lower temperatures.}
\protect\label{fig:12}
\end{figure} 
%_______________________________________________________________

This quantity calculated for our system is reported in fig.\ref{fig:12}.
Also for this parameter we have marked differences with respect to the bulk
that reflect the different weight of the hopping processes in the two cases.
In the bulk the NGP of A and B particles were different and 
never exceeded the values of 2 for A particles (no 
hopping above $T_C$) and 3.2 for B particles. 
Here instead we have similar NGP for A and B particles 
and the value grows enormously as we supercool. For the last temperature
investigated the peak is around 5. We found no relation 
between the peak positions and MCT power law of Eq.\ref{pwl}
as it was found for example in bulk water \cite{fs}
and the curves of the confined NGP do not appear to
follow a master curve  before the maximum 
as found in bulk liquids \cite{kob,fs}.
The trend of the NGP of bulk and confined mixture depicts
a scenario in which MCT behaviour causes a mild deviation from gaussianity.
The peak is around 2 when close to $T_C$~\cite{kob,fs}, its
position shifts substantially
upon supercooling and the NGP follows a master curve in
the temporal region before the peaks. 
Hopping appears to cause a marked deviation from gaussianity instead,
leading to a disappearance both of the MCT 
master curve and of the shift of the
peaks upon supercooling.

\section{Summary and Conclusions}

We presented an MD study of a LJBM confined in a disordered array of
soft spheres. The direct space quantities have been analyzed upon supercooling
for a test of the MCT behaviour.

Two main differences with respect to the bulk 
mixture appear due to confinement.
The $g(r)$ have evidenced that
smaller B particles tend to avoid interfaces at low 
T and correspondingly their
diffusion coefficient becomes lower than that of the A particles.
Hopping processes 
are present also for A particle above $T_C$ as shown
in the VHSCF.

Not obviously, these differences do not prevent 
the MCT asymptotic predictions 
from holding for this mixture also in this kind
of confinement. The range of validity of MCT 
in the region of the $\alpha$ relaxation is however much more limited. 
The range of the bulk \cite{kob} 
is $0.07<\epsilon<1.30$
against the $0.153<\epsilon<0.631$ found in our confined model. 
Disturbances in the dynamical behaviour of the bulk introduced by 
a different type of confinement 
are also reported to lead to
a complete disappearance of MCT~\cite{kobconf}.

Power law fit to the diffusion coefficient 
extracted from the slope of the MSD
give an estimate of $T_C=0.356$ for both 
species and $\gamma=1.86$ for A particles
and $\gamma=1.89$ for B particles in agreement with MCT
that states that these quantities should be the same for both A and B
particles. The values of the bulk extracted from D in literature are \cite{kob}
$T_C=0.435$ for both 
species and $\gamma=2.0$ for A particles
and $\gamma=1.7$ for B particles. The exponents $\gamma$ appear
similar while we have a reduction of $T_C$ upon confinement.
A complete test of the correlators in the Q space 
will be presented in a subsequent paper \cite{noi}.

Acceleration of dynamics in presence of repulsive 
confinement is expected \cite{baschnagel} based on the
fact that in confinement the cooperative 
rearranging region cannot grow beyond the size of the system.
In the case of an attractive interaction this effect 
might superpose
to the often severe slowing down caused by the substrate to the closest 
layers of the liquid\cite{gallo1,gallo2}.

In the present work we have considered a much more complex system
than the bulk. Therefore the fact the MCT could be used in this
context as an unifying theoretical approach is highly relevant
as a guideline for the systematic study of the important
phenomenology of confined and interfacial liquids.

\section {Acknowledgments}

P.G. wishes to thank J. Baschnagel,W. G\"otze and F. Varnik
for very useful discussions.

%%%%%%%%%%%%%%%%%%%%%%%%%%%%%%%%%%%%%%%%%%%%%%%%%%%%%%%%%%%%%%%%%%%%%
\begin{table}
\caption{Parameters of the Lennard-Jones and soft spheres potentials
as defined in Eq.\protect\ref{potential}.
In the Table $A$ and $B$ refer to the particle of the binary mixture
while $M$ refers to the confining soft spheres. Values are expressed
in Lennard-Jones units. \label{tab1}}
 \begin{tabular}{cccc}
    &  $\sigma$  & $\epsilon$ &  $\eta$    \\
 \hline
  $AA$  &  1.0  & 1.0 & 1  \\
  $BB$  &  0.88 & 0.5 & 1  \\
  $AB$  &  0.8  & 1.5 & 1  \\
  $MA$  &  3.0  & 0.32 & 0  \\
  $MB$  &  2.94 & 0.22 & 0  \\
%tableline
 \end{tabular}
 \end{table}
%%%%%%%%%%%%%%%%%%%%%%%%%%%%%%%%%%%%%%%%%%%%%%%%%%%%%%%%%%%%%%%%%%%%%%%%%%%%

\end{document}